\documentclass[%
 reprint,
superscriptaddress,
 amsmath,amssymb,
 aps,
]{revtex4-2}
\usepackage{graphicx}
\usepackage{dcolumn}
\usepackage{bm}

\usepackage{xcolor}

\begin{document}

\preprint{APS/123-QED}

\title{Ultrafast non-thermal suppression of ferroelectricity by carrier screening in LiNbO$_3$}

\newcommand{\MITCHEM}{Department of Chemistry, Massachusetts Institute of Technology, Cambridge, Massachusetts 02139, USA}

\author{Man Tou Wong}
\altaffiliation{These authors contributed equally.}
\author{Zhuquan Zhang}
\altaffiliation{These authors contributed equally.}
\author{Zi-Jie Liu}
\author{Keith A. Nelson}
\email{kanelson@mit.edu}
\affiliation{\MITCHEM}

\begin{abstract}
Ferroelectric materials are key to energy-efficient electronics, memory, and optical applications. While charge carriers typically screen and suppress ferroelectricity, their role under nonequilibrium conditions remains elusive. Here, we use femtosecond laser pulses to liberate trapped carriers in LiNbO$_3$ and track the response using time-resolved second-harmonic generation and stimulated Raman scattering. Even dilute photoexcited carriers induce a rapid yet enduring suppression of polarization and Raman susceptibility. Fluence- and temperature-dependent analyses confirm the suppression is non-thermal and arises from transient carrier screening. These findings reveal an efficient, reversible, and symmetry-preserving mechanism to modulate ferroelectricity on ultrafast timescales, offering a new route to control ferroic and competing quantum phases.

\end{abstract}

\maketitle

Ferroelectrics, characterized by their switchable macroscopic polarization, are of considerable interest due to emergent non-reciprocal responses \cite{tokura2018nonreciprocal,suarez2025nonlinear} and broad applications in energy conversion, memory devices, and information processing \cite{scott2007applications}. Central to these phenomena is the interplay between ferroelectric order and charge carriers. Although free carriers typically screen long-range dipole-dipole interactions and suppress ferroelectricity, the coexistence of carriers and ferroelectric order can produce diverse phenomena, including ferroelectric polarons in semiconductors \cite{miyata2018ferroelectric,franchini2021polarons}, ferroelectric metallicity \cite{anderson1965symmetry,shi2013ferroelectric,bhowal2023polar}, and ferroelectric superconductivity \cite{rischau2017ferroelectric,hameed2022enhanced,jindal2023coupled}. 

Although the impact of carriers on ferroelectricity has been extensively studied under equilibrium conditions \cite{raghavan2016probing,rischau2017ferroelectric,michel_interplay_2021,li2021free,gu2021carrier,jindal2023coupled,li2025unraveling}, photoexcited carriers offer a dynamic means of modulating ferroelectric states on ultrafast timescales. Recent studies have highlighted the complex role of above-bandgap laser excitation in modifying \cite{daranciang2012ultrafast,schick2014localized,jiang2016origin}, redistributing \cite{iwano2017ultrafast,lian2019indirect,lee2021structural,yu2024ultrafast,gao2024large,yang2024light}, or promoting \cite{krapivin2022ultrafast} underlying ferroelectric order. The generation of high-density photocarriers (typically $10^{20}$ cm$^{-3}$ \cite{paillard2019photoinduced}) followed by above-bandgap pumping has been demonstrated as a practical method for ferroelectric polarization screening in ferroelectric and multiferroic materials \cite{takahashi2006terahertz,rana2009understanding,sheu2012ultrafast,zhang2021probing,hoang2025ultrafast,he2016evolution}. However, the resulting polarization dynamics often manifest as a complex relaxation process of hot carriers intertwined with the subsequent photoinduced lattice/spin responses due to electrostriction and inverse piezoelectric effects \cite{chen2012ultrafast,hoang2025ultrafast}. Moreover, such coupled dynamics, mostly studied in engineered thin films or heterostructures, tends to have strong dependence on sample designs and configurations. In contrast to the relatively well-studied high carrier density regime, the possibility of ultrafast ferroelectricity control with a dilute carrier density remains an open question. Despite a limited number of proposals related to this fundamental question (e.g., via below-bandgap excitations \cite{lian2019indirect,chen2024ferroelectric}), experimentally, it remains unclear whether dilute photoinduced carriers can effectively screen polarization—analogous to screening by carriers injected via electrodes—but on femtosecond to picosecond timescales, and how the lattice structure responds to such non-equilibrium carrier populations.

\begin{figure}[b]
\centering
\includegraphics[width = 85 mm]{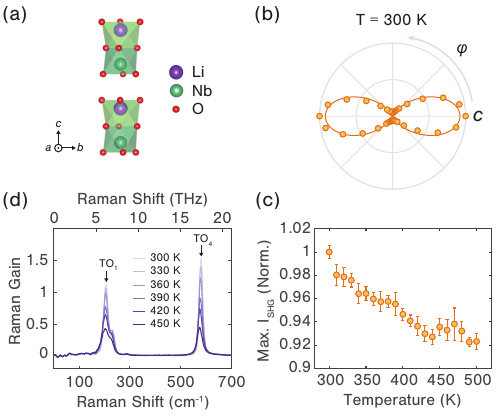}
\caption{\label{fig:1} Crystal structure and equilibrium characterization of LiNbO$_3$. (a) Crystal structure of ferroelectric LiNbO$_3$ in the $R3c$ phase, highlighting the relative displacements of Li and Nb ions with respect to the oxygen octahedra along the hexagonal $c$-axis. (b) Equilibrium SHG response of $x$-cut LiNbO$_3$ in the parallel polarization geometry ($\mathbf{E}_{\mathrm{SHG}} \parallel \mathbf{E}_{\mathrm{probe}}$). The upper panel shows SHG polarimetry at 300 K (solid circles) and the corresponding $|\cos^2\varphi|^2$ fit (solid line). (c) Temperature-dependent SHG amplitudes illustrating the gradual decrease in macroscopic polarization as the temperature is raised toward $T_c$. Standard deviations are indicated. (d) Temperature-dependent stimulated Raman spectra showing the soft mode at 210 cm$^{-1}$ (TO$_1$) and a higher-frequency $A_1$ mode at 591 cm$^{-1}$ (TO$_4$). The decreasing phonon amplitudes reflect the weakening of ferroelectric order with increasing temperature. The suppression of stimulated Raman signals is more pronounced than that of the SHG and is discussed in the main text.}
\end{figure}

\begin{figure*}
\centering
\includegraphics[width = \textwidth]{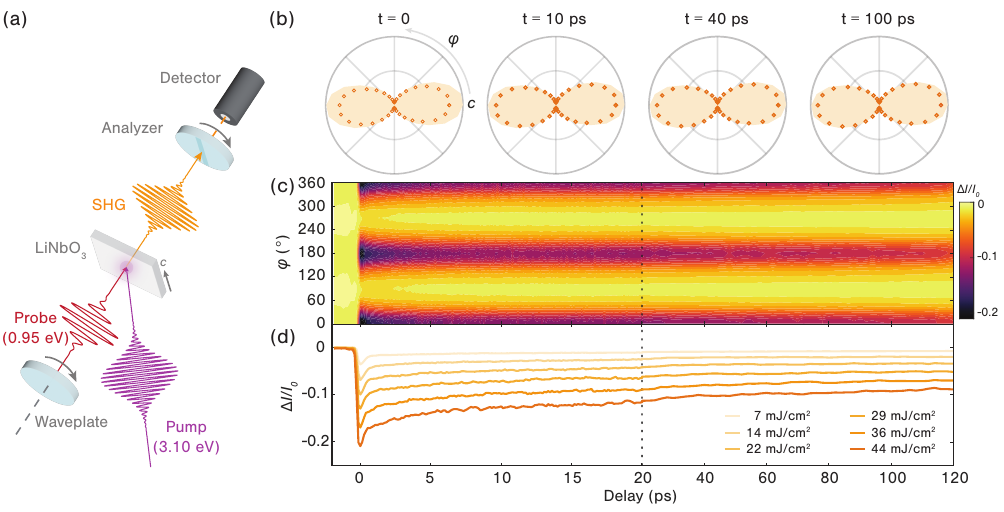}
\caption{\label{fig:2} Time-resolved SHG polarimetry of LiNbO$_3$ at 300 K. (a) Schematic illustration of the time-resolved SHG polarimetry setup. The probe polarization is controlled by a half-wave plate, and SHG signals are detected in the parallel configuration ($\mathbf{E}_{\mathrm{SHG}} \parallel \mathbf{E}_{\mathrm{probe}}$) by rotating a polarizer accordingly. (b) SHG polarimetry at selected pump-probe delays (colored dots), compared to the equilibrium SHG pattern (shaded area). The $|\cos^2\varphi|^2$ dependence is preserved after photoexcitation. (c) Time evolution of normalized SHG amplitudes following excitation with a 44 mJ/cm$^2$ pump. An initial rapid drop of the signal is followed by a slow recovery. (d) Pump-induced SHG dynamics at $\varphi = 0^\circ$ for different pump fluences}
\end{figure*}

In this work, we employ intense, tailored laser pulses to liberate trapped carriers in the prototypical ferroelectric LiNbO$_3$, and we probe the resulting modifications in the ferroelectric order parameter and low-energy collective modes using time-resolved second-harmonic generation (SHG) and Raman scattering, respectively. Our multimodal spectroscopic investigation reveals that even a low density of carriers can efficiently suppress ferroelectricity in an ultrafast, non-thermal manner.

LiNbO$_3$ crystallizes in a rhombohedrally distorted perovskite structure and becomes ferroelectric (space group: $R3c$) below the Curie temperature ($T_c$) around 1480 K \cite{bernhardt_ferroelectric_2024}. In the ferroelectric phase, the Nb and Li ions displace relative to the oxygen octahedra (see Fig. \ref{fig:1}(a)), resulting in a macroscopic polarization along the hexagonal $c$ axis \cite{mankowsky_ultrafast_2017,uzundal2021polarization}. To probe the ferroelectric order, we conducted SHG polarimetry measurements at 300 K by continuously rotating the polarization of a normal-incidence incoming pulse at 0.95 eV ($\mathbf{E}_{\mathrm{probe}}$) and detecting the second-harmonic signal at 1.90 eV ($\mathbf{E}_{\mathrm{SHG}}$) in a parallel configuration ($\mathbf{E}_{\mathrm{SHG}}\parallel \mathbf{E}_{\mathrm{probe}}$). The resulting SHG angular dependence shown in Fig. \ref{fig:1}(b) yields a two-fold symmetric pattern with maxima at $\varphi=0$ (corresponding to $\mathbf{E}_{\mathrm{probe}} \parallel c$), which is consistent with a (noncentrosymmetric) polar point group ($C_{3v}$) and arises from the $d_{33}$ element of the second-order susceptibility \cite{mankowsky_ultrafast_2017,zu_analytical_2022}. As the temperature increases, the equilibrium SHG intensity decreases, reflecting the gradual reduction of macroscopic polarization as the system approaches $T_c$, further confirming the sensitivity of the SHG signal to the ferroelectric order parameter (Fig. \ref{fig:1}(c)). 

Complementary insights into lattice dynamics associated with ferroelectricity are provided by Raman scattering. As shown in Fig.~\ref{fig:1}(d), the stimulated Raman spectra reveal two prominent phonon modes at 210 (TO$_1$) and 591 cm$^{-1}$ (TO$_4$), attributed to the ferroelectric soft mode and another $c$-axis-polarized $A_1$ mode, respectively \cite{schaufele_raman_1966,barker_dielectric_1967,servoin_soft_1979,sanna2015raman,henstridge_nonlocal_2022}. Both modes exhibit a systematic decrease in amplitude with increasing temperature, further indicating the weakening of ferroelectric order as the system is brought closer to $T_c$ \cite{sakamoto_temperature_1976,ridah_temperature_1997}. Compared with the SHG results in Fig.~\ref{fig:1}(c), the suppression of the stimulated Raman signals is more pronounced, which resembles the drastic decrease in spontaneous Raman scattering intensities in LiTaO$_3$ and LiNbO$_3$ at high temperatures \cite{johnston1968temperature,ridah_temperature_1997}. Previous experimental studies showed that the $d_{ij}$ values of LiNbO$_3$ have weak temperature dependences in the 300-500 K range \cite{miller1966temperature,bergman1976molecular}, and our SHG results agree with this trend reported in literature. Since the temperature-induced change in Raman responses depends on both electronic (e.g., Raman tensor elements) and phonon (e.g., thermal population and dephasing) contributions \cite{batignani2024temperature}, their combined effects may account for the stronger temperature sensitivity observed in Raman than in SHG, which is determined by electronic nonlinearity.

Having characterized the equilibrium properties of the ferroelectric phase in LiNbO$_3$ via SHG and Raman scattering, we next investigated the influence of photoexcited carriers on ultrafast timescales. To this end, we applied pump pulses centered at 3.10 eV to release trapped electrons from oxygen vacancy defect states located below the band gap ($E_g = 3.78$ eV) \cite{sweeney_oxygen_1983,dhar_optical_1990}, at a sample temperature of 300 K. We first examined the impact of these free carriers on the ferroelectric order parameter through time-resolved SHG polarimetry measurements (Fig.~\ref{fig:2}(a)). In our setup, the normal-incidence probe beam induces the SHG signal detected in the same configuration as equilibrium measurement, i.e., $\mathbf{E}_{\mathrm{SHG}} \parallel \mathbf{E}_{\mathrm{probe}}$. The pump-induced changes in SHG intensity were recorded as a function of the pump-probe delay.

Figures~\ref{fig:2}(b) and \ref{fig:2}(c) show the time evolution of the normalized SHG polarimetry. The SHG intensity exhibits a rapid, sub-picosecond drop immediately after photoexcitation, followed by a slow recovery over timescales exceeding 100 ps (see Supplemental Material Note. 5). Notably, the initial suppression reaches approximately 20\% at the highest pump fluence. This prompt reduction reflects an ultrafast suppression of the macroscopic polarization and suggests efficient screening of the ferroelectric order by photoinduced carriers. Despite the amplitude change, the SHG polarimetry maintains its equilibrium $|\cos^2\varphi|^2$ angular dependence (Fig.\ref{fig:2}(b)), indicating that the crystal symmetry remains unaltered. The slow recovery dynamics are consistent with the timescales of carrier trapping and recombination, as previously reported for LiNbO$_3$ \cite{beyer_investigation_2005}.

The pump fluence dependence measurements (Fig. \ref{fig:2}(d)) show that the recovery dynamics of the SHG signals at different fluence values follow the same trend, which implies that the ferroelectricity reduction mechanism is the same within the pump fluence range that we explored in this work. The comparison between time-resolved and equilibrium (Fig. \ref{fig:1}(b)) SHG experiments, particularly in the suppression of the maximum SHG intensity, will be discussed later.

\begin{figure}
\centering
\includegraphics[width = 85 mm]{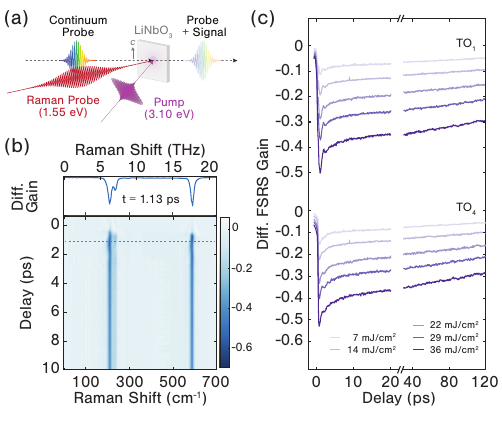}
\caption{\label{fig:3} Ultrafast Raman response of LiNbO$_3$ probed by time-resolved FSRS. (a) Schematic illustration of the time-resolved FSRS setup. The Raman- and continuum-probe pulses are polarized along the $c$ axis to selectively excite and detect $A_1$ phonon modes. (b) Time evolution of FSRS signals following excitation with a 36 mJ/cm$^2$ pump. The upper panel shows the FSRS spectrum at the delay corresponding to the strongest suppression. Both phonon modes show significant reductions in differential Raman gain. (c) Transient FSRS traces of TO$_1$ and TO$_4$ modes for various pump fluences, showing prolonged suppression consistent with photoinduced carrier dynamics.}
\end{figure}

\begin{figure}
\centering
\includegraphics[width = 85 mm]{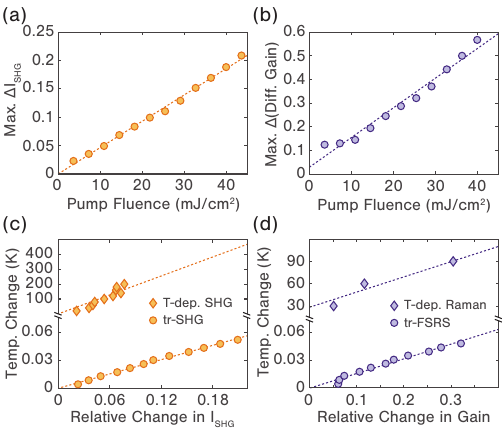}
\caption{\label{fig:4} Fluence and temperature dependence of ferroelectric suppression in LiNbO$_3$.(a) Maximum suppression of SHG signal at $\varphi = 0^\circ$ as a function of pump fluence, showing a linear response.
(b) Maximum suppression of FSRS differential gain of TO$_1$ versus pump fluence, also displaying linear behavior.
(c) Comparison between pump-induced and temperature-induced changes in SHG intensity, with fluence converted to effective temperature using the Debye model.
(d) Similar comparison for Raman gain at TO$_1$. In both (c) and (d), the pump-induced effects correspond to much smaller effective temperature rises than those required under thermal equilibrium, confirming the non-thermal nature of the ferroelectricity suppression.}
\end{figure}

To investigate how the lattice structure responds to free-carrier generation, we performed time-resolved femtosecond stimulated Raman scattering (FSRS) experiments, schematically depicted in Fig. \ref{fig:3}(a). In time-resolved FSRS \cite{lee_theory_2004,dietze_femtosecond_2016,batignani_femtosecond_2024}, the non-equilibrium dynamics initiated by the pump pulse (3.10 eV) are probed via third-order light-matter interactions involving a narrowband Raman probe pulse (1.55 eV) and a continuum probe pulse. The resulting FSRS signals are emitted along the same direction as the continuum probe pulse, which are both detected by the spectrometer to yield the heterodyned Raman peaks. In our measurements, both Raman- and continuum-probe fields were polarized along the $c$ axis of LiNbO$_3$ to exclusively probe the Raman phonon modes with $A_1$ symmetry. The differential Raman gain, defined as the pump-induced change of the FSRS signal, is extracted and plotted as a function of pump-probe delay for each of the two modes in Fig. ~\ref{fig:3}(b).

The time-resolved FSRS signal reveals that the liberated carriers have a pronounced and long-lived effect on the Raman response. Following initial coherent ripples before time zero—attributed to electro-optic artifacts \cite{peterson_electro-optic_1964,kaminow_quantitative_1967}—the differential gain drops sharply (by 50\%) across both phonon bands within the first 2 ps. This pronounced suppression suggests substantial modifications of the Raman susceptibility tensor. The absence of new phonon bands and negligible frequency shift of the original phonon bands in the FSRS spectra implies that the pump-induced changes are predominantly electronic rather than structural. The prolonged suppression of Raman signals across various pump fluences (Fig. \ref{fig:3}(c)) mirrors the pump-induced SHG reduction. These observations indicate that both responses originate from carrier screening with a long relaxation time.

The pump fluence-dependent FSRS signal suppression exhibits a trend similar to that seen in equilibrium temperature-dependent FSRS spectra, where the Raman gain diminishes as the system is brought closer to $T_c$ (Fig.~\ref{fig:1}(d)) \cite{sakamoto_temperature_1976,ridah_temperature_1997}. This parallel suggests a link between Raman signal reduction and ferroelectric suppression. Since carrier de-trapping can also be thermally induced, and the Raman susceptibility is sensitive to local field fluctuations \cite{sakamoto_temperature_1976}, the observed behavior is consistent with a carrier screening mechanism, where charge carriers screen the Coulomb interactions that favor the off-center displacements and partially quench ferroelectricity. To rigorously distinguish thermal contributions from non-thermal effects, we systematically compared pump-fluence-dependent SHG and FSRS results with their temperature-dependent counterparts.

Figures~\ref{fig:4}(a) and \ref{fig:4}(b) show the pump fluence dependence of the maximum amplitude changes in time-resolved SHG and time-resolved FSRS signals, respectively. In both cases, the response scales linearly with the pump fluence, indicating that the ferroelectric suppression originates from liberating trapped electrons via one-photon absorption. As a comparison, subgap excitation does not result in a long-lived FSRS signal (see Supplemental Material Note. 6). To assess whether the observed effects are thermal in nature, we estimated the effective temperature rise using a Debye model \cite{mann_probing_2016,gao_snapshots_2022} (see Supplemental Material Note. 7), and compared the amplitude changes to those in equilibrium SHG and FSRS experiments.

Figures~\ref{fig:4}(c) and \ref{fig:4}(d) compare temperature- and pump-induced changes in SHG intensity and Raman gain of the soft mode, respectively. Achieving comparable signal suppression via laser excitation requires an effective temperature rise orders of magnitude smaller than that needed under thermal equilibrium. This discrepancy clearly demonstrates the non-thermal nature of the ferroelectricity suppression.

In addition to minimal pump-induced heating, the estimated upper bound of the photoexcited carrier density responsible for the observed signal changes is exceedingly low—on the order of $10^{-6}\ e$ per unit cell (see Supplemental Material Note. 8). Crucially, because the pump photon energy lies below the band gap, the optical penetration depth is exceptionally large ($\sim12.7$ mm), enabling carrier-mediated modulation of polarization throughout the bulk of the crystal. This stands in sharp contrast to above-bandgap excitation, where strong absorption confines carrier generation—and thereby polarization screening—to a near-surface region typically limited to submicron length scales. While hot carriers may diffuse modestly beyond the initial absorption volume, such transport is neither spatially uniform nor readily controllable, and is unlikely to extend beyond $\sim1~\mu$m. Our results therefore demonstrate that below-bandgap excitation provides a fundamentally different, bulk-sensitive route to ultrafast ferroelectric control. The carrier density generated by the present below-bandgap excitation protocol, $10^{16}–10^{17}$ cm$^{-3}$, is substantially lower than the local carrier densities typically invoked in electrode-injection studies of polarization reversal or strong polarization modification, which are commonly in the range of $10^{18}–10^{22}$ cm$^{-3}$\cite{buhlmann_polarization_2005,rault2014reversible}. We emphasize, however, that these two situations are not directly equivalent: carriers injected from electrodes are usually concentrated near ferroelectric/conductive interfaces, whereas the subgap optical excitation used here generates a dilute carrier population throughout the bulk crystal. This spatially uniform bulk screening distinguishes the present approach from both above-bandgap photoexcitation and conventional carrier injection. 

Together, the time-resolved SHG and FSRS results reveal persistent signal suppression following ultrafast carrier generation, indicating a competing interaction between charge carriers and macroscopic polarization in ferroelectric materials. The fluence dependence of the signal dynamics suggests that this reduction is tunable, offering a pathway for dynamically controlling ferroelectric states in an ultrafast, non-thermal manner. Noted that the observed transient weakening of polar order parameter in LiNbO$_3$ is induced by the photocarriers liberated from oxygen-vacancy defects distributed in the bulk, which utilizes defects as an intrinsic charge reservoir and is distinct from the conventional defect-driven destabilization of equilibrium ferroelectric phase via chemical or charge doping. Rather than proposing a new microscopic carrier-ferroelectricity interaction different from carrier screening, we identify a below-bandgap excitation limit that differs mechanistically  from conventional above-gap pumping, where carrier generation is typically surface-localized and coupled to lattice strain responses. Investigating other ferroelectric materials and developing quantitative microscopic models for mechanistic insights remain important future directions, which fall beyond the scope of the present work. In contrast to electrostatic control of ferroelectricity explored in previous theoretical and experimental works \cite{grekov_encountering_1976,buhlmann_polarization_2005,maksymovych_polarization_2009,wang_ferroelectric_2012,maksymovych_tunable_2012,crassous_polarization_2015,zhu_tuning_2020,michel_interplay_2021,jindal2023coupled}, the carrier-induced effects demonstrated here unfold on sub-picosecond timescales and produce long-lived responses. Such an efficient, reversible, and non-thermal suppression mechanism not only opens new possibilities for the ultrafast manipulation of ferroelectric materials, especially those with competing or intertwined orders\cite{shi2013ferroelectric,zheng2020unconventional,song2022evidence,jindal2023coupled,dastrup2025electromagnon,bustamante2025ultrafast}, but also motivates methodological development using the potentially advantageous perturbative carrier generation enabled by below-bandgap excitation to study domain-wall dynamics and directional carrier transport in quantum materials.

\begin{acknowledgments}
This work was supported by the U.S. Department of Energy, Office of Basic Energy Sciences, under Grant No. DESC0019126. This work was carried out in part through the use of MIT Materials Research Laboratory's facilities.
\end{acknowledgments}

\bibliography{main}

\end{document}